\documentclass[review,number,sort&compress,3p]{elsarticle}
\usepackage{lineno,hyperref}

\journal{Chinese Physics C}

\usepackage{amsmath}
\usepackage{amsfonts}
\usepackage{amssymb}
\usepackage{graphicx}
\usepackage{multirow}
\usepackage{booktabs}
\usepackage{soul,color}
\setstcolor{red}
\usepackage{bm}
\usepackage{subcaption}
\usepackage{makecell}
\usepackage{footnote}

\begin{document}
\graphicspath{{Figure/}}

\begin{frontmatter}

\title{Muon Flux Measurement at China Jinping Underground Laboratory}


\author[THU,HEP]{Ziyi Guo}
\author[THU,TBU]{Lars Bathe-Peters\fnref{HU}}
\author[THU,HEP,LAB]{Shaomin Chen}
\author[THU,EPFL]{Mourad Chouaki}
\author[THU,HEP]{Wei Dou}
\author[THU,HEP]{Lei Guo}
\author[THU,HEP]{Ghulam Hussain\fnref{BUIT}}
\author[THU,HEP]{Jinjing Li}
\author[UCAS]{Qian Liu}
\author[SYSU]{Guang Luo}
\author[UCAS]{Wentai Luo}
\author[NJU]{Ming Qi}
\author[THU,HEP]{Wenhui Shao}
\author[SYSU]{Jian Tang}
\author[THU,HEP]{Linyan Wan\fnref{BU}}
\author[THU,HEP,LAB]{Zhe Wang}
\author[THU,HEP,LAB]{Benda Xu}
\author[THU,HEP]{Tong Xu}
\author[THU,HEP]{Weiran Xu\fnref{MIT}}
\author[THU,HEP]{Yuzi Yang}
\author[BNL]{Minfang Yeh}
\author[THU,HEP]{Lin Zhao}

\address{}

\author{(JNE Collaboration)}

\fntext[HU]{Now at: Department of Physics, Harvard University, Cambridge, MA 02138, USA.}
\fntext[BUIT]{Now at: Balochistan University of Information Technology, 
Engineering and Management Sciences, Quetta 1800, Pakistan.}
\fntext[BU]{Now at: Boston University, Boston, MA 02215, USA.}
\fntext[MIT]{Now at: Massachusetts Institute of Technology, Cambridge, MA 02139, USA.}

\address[THU]{Department of Engineering Physics, Tsinghua University, Beijing 100084, China}
\address[HEP]{Center for High Energy Physics, Tsinghua University, Beijing 100084, China}
\address[LAB]{Key Laboratory of Particle \& Radiation Imaging (Tsinghua University), Ministry of Education, China}
\address[TBU]{Institut f{\"u}r Physik, Technische Universit{\"a}t Berlin, Berlin 10623, Germany}
\address[EPFL]{{\'E}cole Polytechnique F{\'e}d{\'e}rale de Lausanne, Lausanne 1015, Switzerland}
\address[UCAS]{School of Physical Sciences, University of Chinese Academy of Sciences, Beijing 100049, China}
\address[SYSU]{School of Physics, Sun Yat-Sen University, Guangzhou 510275, China}
\address[NJU]{School of Physics, Nanjing University, Nanjing 210093, China}
\address[BNL]{Brookhaven National Laboratory, Upton, New York 11973, USA}

\begin{abstract}

China Jinping Underground Laboratory (CJPL) is ideal for studying solar-, geo-, and supernova neutrinos. A precise measurement of the cosmic-ray background would play an essential role in proceeding with the R\&D research for these MeV-scale neutrino experiments. Using a 1-ton prototype detector for the Jinping Neutrino Experiment (JNE), we detected 264 high-energy muon events from a 645.2-day dataset at the first phase of CJPL (CJPL-I), reconstructed their directions, and measured the cosmic-ray muon flux to be $(3.53\pm0.22_{\text{stat.}}\pm0.07_{\text{sys.}})\times10^{-10}$\,cm$^{-2}$s$^{-1}$. The observed angular distributions indicate the leakage of cosmic-ray muon background and agree with the simulation accounting for Jinping mountain's terrain. A survey of muon fluxes at different laboratory locations situated under mountains and below mine shaft indicated that the former is generally a factor of $(4\pm2)$ larger than the latter with the same vertical overburden. This study provides a convenient back-of-the-envelope estimation for muon flux of an underground experiment.
\end{abstract}


\end{frontmatter}

\newcommand{\ud}{\mathrm{d}}
\newcommand{\ui}{\mathrm{i}}
\newcommand{\ue}{\mathrm{e}}

\section{Introduction}
\label{sec:intro}

The China Jinping Underground Laboratory (CJPL), located in Sichuan Province, China, is one of the world's deepest underground laboratories~\cite{cjp2017}. The rock overburden at CJPL is about 2400\,m vertically~\cite{kang2010status} and the closest nuclear power plant is approximately 1000\,km away. It is an ideal site for rare-event experiments such as dark matter search~\cite{cdex,pandax}, neutrinoless double beta decay~\cite{cdex2,pandax3}, and solar neutrino study. The proposed Jinping Neutrino Experiment aims to study MeV-scale low-energy neutrinos, including solar neutrinos, geoneutrinos, and supernova relic neutrinos (also referred to as the diffuse supernova neutrino background)~\cite{beacom2017physics,Geo-WanLinyan,Geo-Bill}. 

These studies are very prone to the contamination from cosmic-ray muon and muon-induced radioactive isotope backgrounds. From the dominant vertical muons detected by a plastic scintillator telescope, the first measurement of cosmic-ray flux was $(2.0\pm0.4)\times10^{-10}$\,cm$^{-2}$s$^{-1}$ and had no angular correction to the detector acceptance~\cite{Wu_2013}. Ref.~\cite{Mayor_2015} categorized underground laboratories into two types: below mountains (mountain shape overburden) and down mine shafts (flat overburden). However, as Ref.~\cite{Mei:2005gm} pointed out, the flux magnitude is quite different in different laboratory locations situated under mountains and below mine shafts with the same vertical rock overburden. This difference can lead to different background levels for a variety of physics implications, such as the cosmogenic $^{11}$C background in search for Carbon-Oxygen-Nitrogen solar neutrinos~\cite{BorexinoCNO}, and the cosmic-ray spallation background in searches for the upturn of the solar $^{8}$B neutrino spectrum~\cite{sks} and supernova relic neutrinos~\cite{PhysRevD.93.012004}. Therefore, a precise total flux measurement and a detailed cosmic-ray leakage study are necessary for the active and passive shielding design of future neutrino experiments. 

A 1-ton scintillator detector serves as a prototype of the Jinping Neutrino Experiment  and has been running since 2017~\cite{1ton}. This prototype aims to test the performance of several related key detector components, understand the neutrino detection technology, and measure the underground background level in situ. This study used this omnidirectional detector to measure the cosmic-ray muon flux at CJPL-I, including the muon angular distributions, which enable a clear understanding of the cosmic-ray leak through the mountain topography profile. 

After detailing the design of the 1-ton prototype detector, we describe the model for predicting the underground muon energy and angular distributions, muon event selection, and direction reconstruction. In the end, the muon flux measurement based on the two-year data of the 1-ton prototype is reported in this study.

\section{The 1-ton prototype detector}
\label{sec:det}

\begin{figure}[htbp]
\centering
\includegraphics[width=0.6\textwidth]{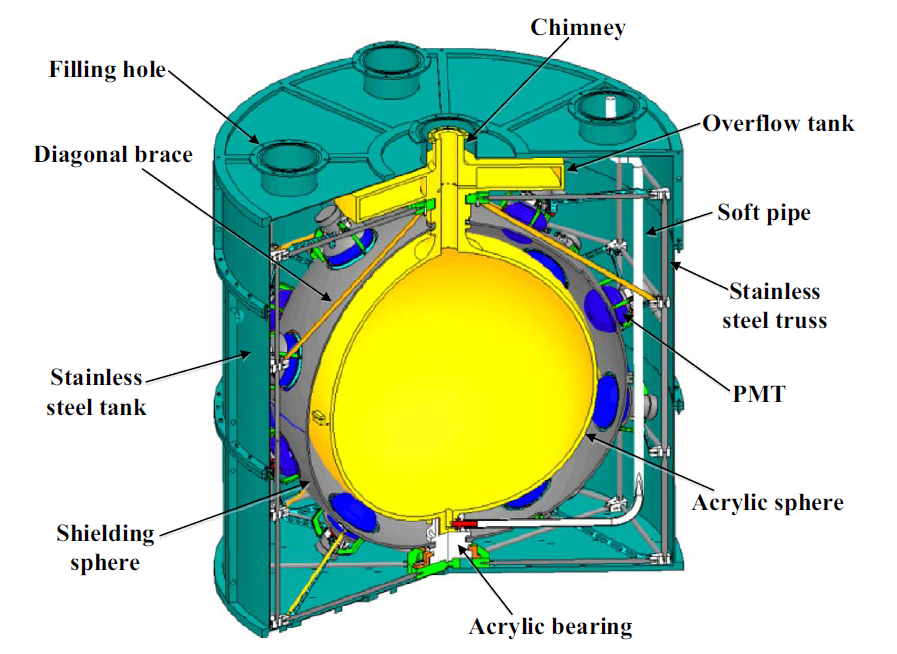}
\caption{1-ton prototype of Jinping Neutrino Experiment.}
\label{fig:1ton}
\end{figure}

Figure~\ref{fig:1ton} shows the detector's schematic structural diagram. To reduce the environmental background, we used 20\,cm$\times$10\,cm$\times$5\,cm lead bricks to form a shielding wall outside the tank (not drawn in the figure). The detector measures 2\,m height and contains one ton of custom liquid scintillator in a 0.645\,m-radius acrylic spherical vessel~\cite{1ton}. This scintillator, referred to as the slow liquid scintillator~\cite{li2016separation,sls}, is a linear alkylbenzene (LAB) doped with 0.07\,g/l of the fluor 2,5-diphenyloxazole (PPO) and 13\,mg/l of the wavelength shifter 1,4-bis (2-methylstyryl)-benzene (bis-MSB). This slow scintillator delays the scintillation light emission duration, thus enhances the Cherenkov-to-scintillation light ratio in the early arrival time to separate these two lights in high efficacy. Thirty 8-inch Hamamatsu R5912 photomultiplier tubes (PMTs) outside the acrylic vessel detected the Cherenkov and scintillation lights and output their pulse shapes to front-end electronics. A water buffer layer between the outer layer of the acrylic vessel and the inner wall of the stainless steel tank serves as a passive shielding material to suppress the ambient radioactive background. 

The front-end electronic system included 4 CAEN V1751 FlashADC boards and one logical trigger module CAEN V1495. Each FlashADC board had eight channels, 10 bit ADC precision for 1\,V dynamic range, and 1\,GHz sampling rate. All the PMT signals directly went into V1751 for digitization. If more than 25 PMTs got fired, the data acquisition system would record all the fired PMTs' pulse shapes in a 1029-ns time window.

\section{The predicted muon energy spectrum and angular distribution}
\label{sec:mpre}

The energy spectrum and angular distribution of underground muons were used as the inputs for the detector simulation. The muon direction is defined as along where it comes from throughout this paper. A Geant4~\cite{G41, G42}-based package simulated muon penetration development in the mountain rock to predict various underground muon characteristic profiles, with its own standard electromagnetic and muon-nucleus processes. 

\begin{figure}[htbp]
\centering    
\includegraphics[width=0.4\textwidth]{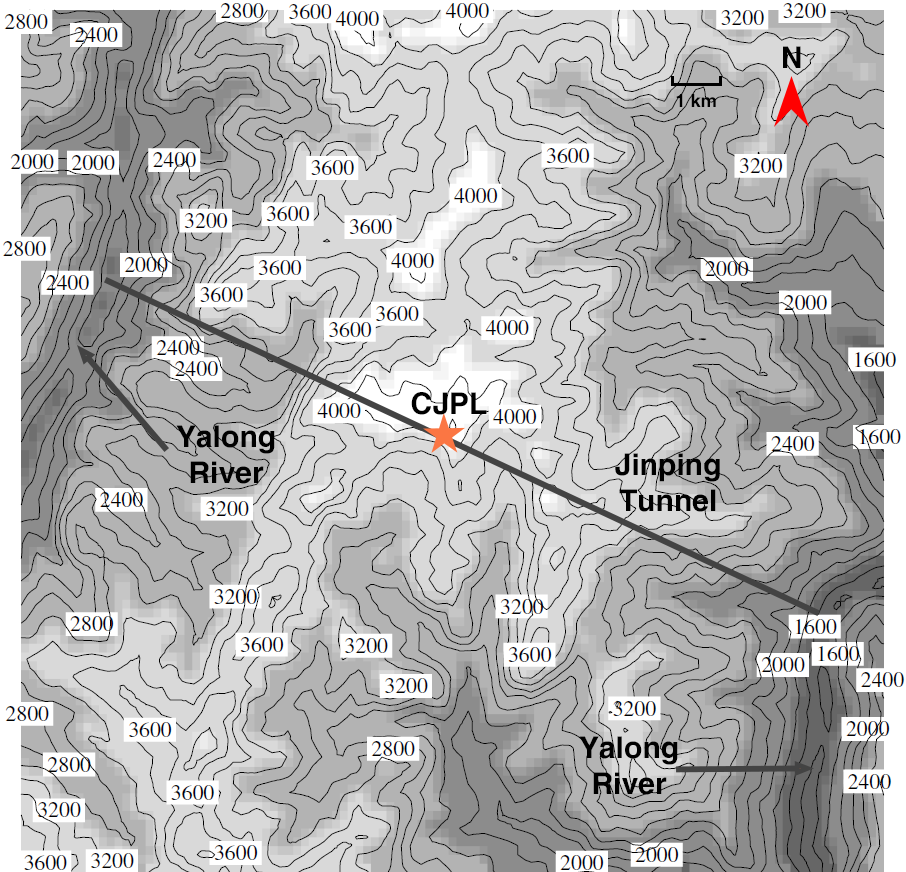}
\caption{The contour map near CJPL-I, as given by the SRTM3 dataset~\cite{nasa}.}    
\label{fig:terr}
\end{figure}

Jinping mountain is about 4000\,m above sea level, and the elevation of the experimental hall is about 1600\,m. We obtained the mountain terrain data from the NASA SRTM3 dataset~\cite{nasa}. Figure~\ref{fig:terr} shows the contour map. There were 6315 survey points within a 9\,km radius circle centered at the laboratory. We assembled them to a mesh using Delaunay triangulation, a standard algorithm, to divide discrete points into a set of triangles with the restriction that two adjacent triangles entirely share with each triangle side.

We assumed Jinping mountain's average rock density to be 2.8\,g/cm$^3$ from Ref.~\cite{Wu_2013}, so the water equivalent depth was 6720\,m for 2400\,m rock. The density variation can affect the simulated spectrum. However, it is negligible for the flux measurement since, in our case, the detection efficiency is not sensitive to the spectrum, as discussed in Sec.~\ref{sec:unc}.  The composition of the rock in the simulation utilized the values from the abundance of elements in Earth’s crust (percentage by weight)\cite{crc}: oxygen (46.1\%), silicon (28.2\%), aluminum (8.2\%), and iron (5.6\%). The modified Gaisser’s formula~\cite{guan} parametrized cosmic-ray muon's kinetic energy $E$ and zenith angle $\theta$ distribution at sea level below,

\begin{equation}
\label{eq:mg}
\begin{aligned}
&G(E,\theta)\equiv\frac{\ud N}{\ud E \ud \Omega}=\\
&\frac{I_0}{\text{cm}^2\cdot \text{s}\cdot \text{sr}\cdot\text{GeV}}\cdot \left(\frac{E^{\star}}{\text{GeV}}\right)^{-\gamma}\cdot
\left(\frac{1}{1+\dfrac{1.1E\cos\theta^\star}{115\text{GeV}}}+\frac{0.054}{1+\dfrac{1.1E\cos\theta^\star}{850\text{GeV}}}\right)
\end{aligned}
\end{equation}
where $E^\star$ and $\cos\theta^\star$ are defined as follow,
\begin{equation}
E^{\star}=E\left[1+\frac{3.64\,\text{GeV}}{E\cdot(\cos\theta^\star)^{1.29}}\right],
\quad
\cos\theta^\star=\sqrt{\frac{\cos^2\theta+P_1^2+P_2(\cos\theta)^{P_3}+P_4(\cos\theta)^{P_5}}{1+P_1^2+P_2+P_4}}
\end{equation}
where $I_0$ is a normalization constant, $\gamma=2.7$ is the muon spectral index, $P_1,P_2,P_3,P_4$, and $P_5$ are parameters in Ref.~\cite{guan}. 

\begin{figure}[htbp]
\centering
\includegraphics[width=0.4\textwidth]{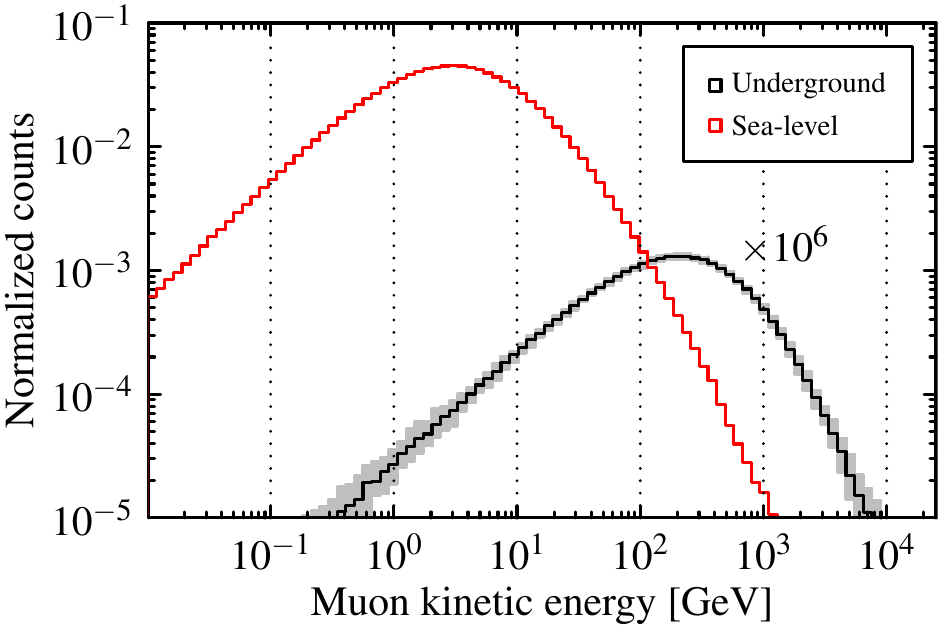}
\caption{Simulation result of underground muon kinetic energy. The mean value is 340\,GeV. The gray band shows the 1$\sigma$ uncertainty. See more details in the text. The spectrum of muons at sea-level is also plotted.}
\label{fig:muonekdist}
\end{figure}

\begin{figure}[htbp]
\centering    
\includegraphics[width=0.65\textwidth]{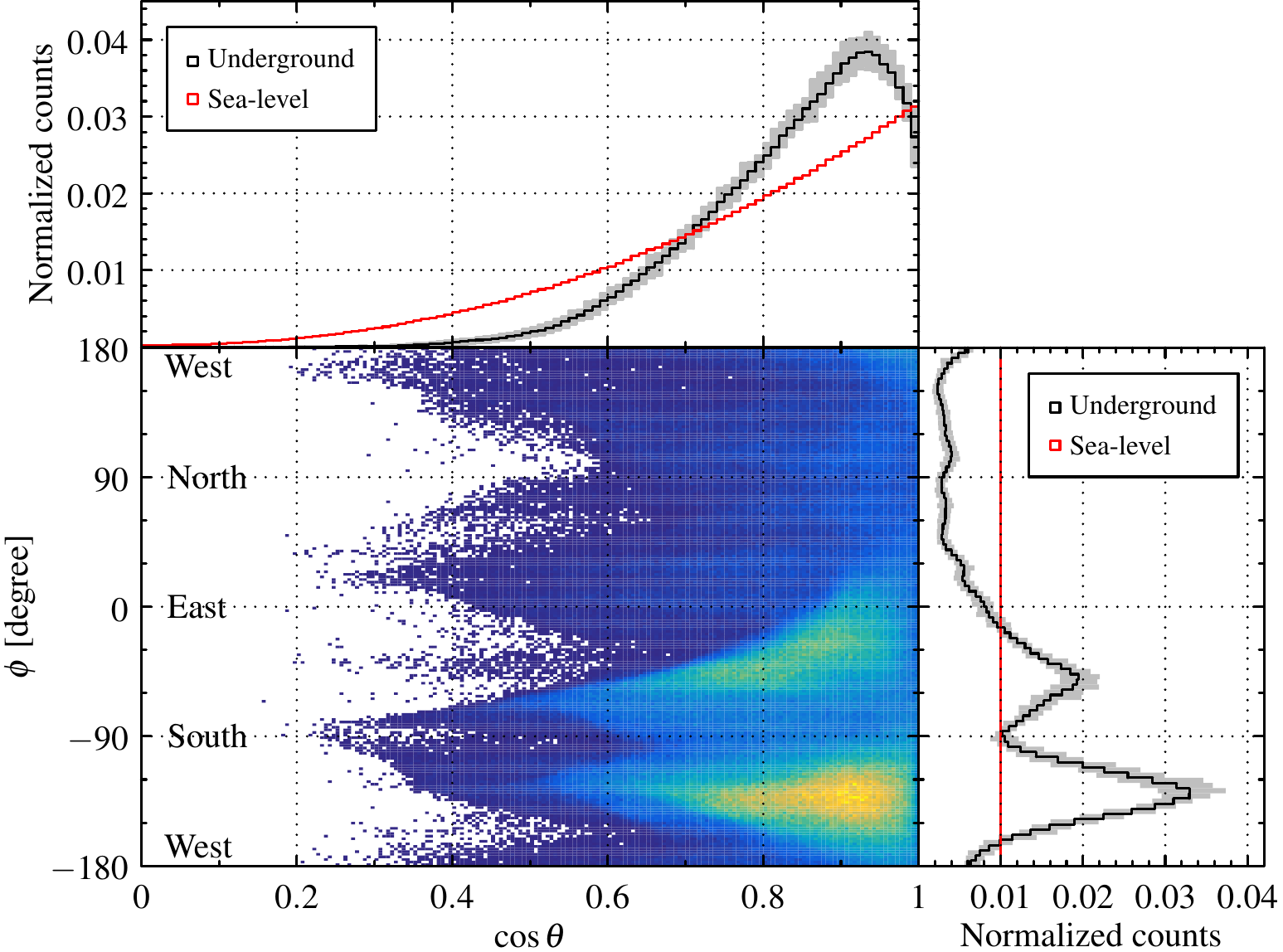}
\caption{Simulation result of underground muon direction $(\cos\theta,\phi)$ and one-dimensional projections. Muons from the south is intensive, as expected from the contour map in Figure~\ref{fig:terr}. The gray band shows the 1$\sigma$ uncertainty. See more details in the text. The spectrum of muons at sea-level is also plotted.}    
\label{fig:muondist}
\end{figure}

Figures~\ref{fig:muonekdist} and \ref{fig:muondist} show the simulated underground muon kinetic energy and corresponding angular distributions at CJPL-I. The uncertainties came from the precision of the NASA’s dataset (90\,m in horizon directions) and the experimental hall size $\sim100$\,m. We also plot the distributions at sea-level for comparison. The expected corresponding zenith angle follows a $\cos^2\theta$ distribution and the azimuth angle follows a uniform distribution. The observed cosmic-ray leak in the south direction agrees with Figure~\ref{fig:terr}, in which the contour plot has already indicated less rock coverage.

Due to the high elevation ($\sim$4000\,m), the altitude and latitude may affect the muon distributions described in Eq.~(\ref{eq:mg}). However, Ref.\cite{Cecchini1998kv} pointed out that the differential flux at high energy ($>40$\, GeV) and small zenith angle barely depends on altitude and latitude. Since the minimum energy required for muons to reach CJPL-I is approximately 3\,TeV, and the cosine of the zenith angle concentrates above 0.4, we concluded that CJPL-I's altitude and latitude do not affect the underground muon spectrum.

\section{Event selection}
\label{sec:evt}

This study analyzed the data collected from July 31, 2017 to July 12, 2019. We first required that runs should be flagged as good runs, i.e., neither pedestal calibration nor detector maintenance. Data quality check parameters for identifying apparent noise were the trigger rate, baseline, and baseline fluctuation of a waveform. A data file should not have these quantities deviated from the reference values by three standard deviations. The live time after data qualify check was $5.575\times10^7$\,s, or 645.2 live days.

We then required a minimum number of photoelectrons (PEs), corresponding to approximately 100\,MeV energy deposits or 50\,cm track length in the scintillator. When passing through the detector's edge, a muon deposits less energy and is indistinguishable with that from the radioactive background, muon shower, or noise events. Therefore, this cut discarded low-energy events to get a high purity sample.

We finally removed the electronics noise and flasher events, which were highly-charged light-emitting events, possibly from PMT bases' discharging. Examining all the high energy deposit events' waveforms, we found that some of them always had a single PMT with a much higher charge than the others, while a muon event was of a more uniform charge distribution. We defined a ratio of maximum PE number of each PMT to total PE number in one event, notated as $r_{\text{max}}$, should not be greater than 0.15 to identify the flasher events. 

\begin{figure}[htbp]
\centering
\includegraphics[width=0.7\textwidth]{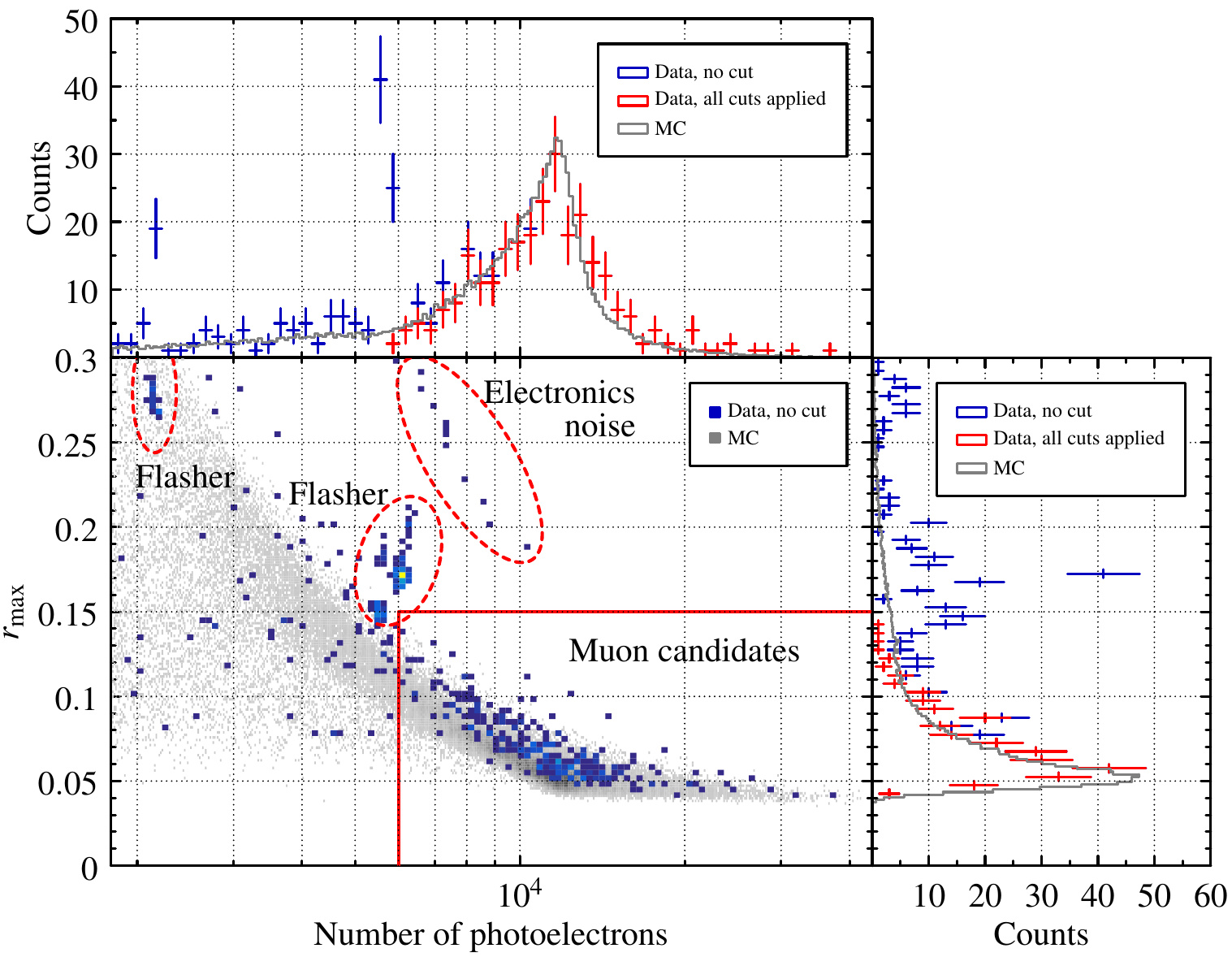}
\caption{The scattered plot and one-dimensional projections of $r_{\text{max}}$ and PE number distribution from the data. The grey area in the two-dimensional distribution is the simulation result. Typical muon candidates spread in the region of $r_{\text{max}}<0.15$ and PE number $>6000$, while flasher and electronics noise events have larger $r_{\text{max}}$ and distribute in some clusters marked with circles. Low-energy events (PE number $<6000$) may contain indistinguishable radioactive background, shower, or noise events are also be removed.}
\label{fig:Ekmax} 
\end{figure}

Figure~\ref{fig:Ekmax} shows a two-dimensional distribution and one-dimensional projections of PE number and $r_{\text{max}}$, indicating that the flasher events and the electronic noise events correspond to the clusters with larger $r_{\text{max}}$. We also plotted the simulation result and one-dimensional projections for better comparison. In the end, 264 muon candidates passed the selection criteria. Table~\ref{tab:cuts1} summarizes all Selection criteria for muon candidates selection.

\begin{table}[htbp]
\centering
\caption{Summary of cuts for muon candidates selection.}
\label{tab:cuts1}
\begin{tabular}{cc}
\toprule
\noalign{\smallskip}
Type  & Selection criteria \\ 
\noalign{\smallskip}
\midrule
\multirow{2}{*}{\makecell[c]{Data quality\\check}}  & Good run\\ 
  & Trigger rate, baseline and baseline fluctuation \\ 
\midrule
\multirow{2}{*}{\makecell[c]{Muon candidates\\selection}} &  Number of photoelectrons $>6000$  \\ 
 & $r_{\text{max}}<0.15$ \\ 
\bottomrule
\end{tabular} 
\end{table}

\section{Direction reconstruction}
\label{sec:rec}

\begin{figure}[htbp]
\centering
\includegraphics[width=0.3\textwidth]{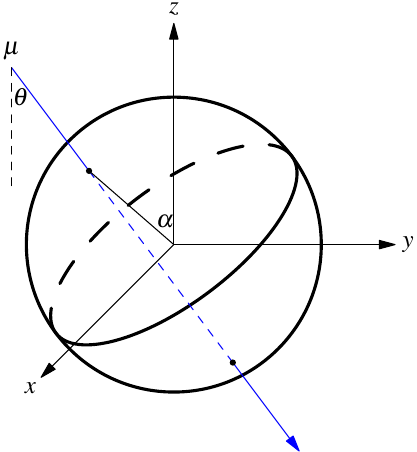}
\caption{Muon generator in the PMT trigger time pattern template. The muon direction $(\cos\theta,\phi)$ and entry point $(\cos\alpha,\beta)$ were sampled uniformly.}
\label{fig:recon} 
\end{figure}

We used a template-based method to reconstruct the muon direction. The templates were generated from a Geant4-based simulation. Each template was tagged with the muon direction $\mathbf{p}_i=(\cos\theta,\phi)$ and the entry point on the acrylic vessel $(\cos\alpha,\beta)$, as shown in Figure~\ref{fig:recon}. 
When a muon's direction was sampled from a uniform distribution, its entry point on the vessel surface was also sampled uniformly on the hemisphere facing the muon direction.

About 250k template events passed the event selection criteria described in Section~\ref{sec:evt}. For the PMT arrival time pattern vector of template $i$: $\mathbf{T}_i=(t_{0i},t_{1i},\cdots,t_{29i})$, we subtracted the mean value $\bar{t}_i=\frac{1}{30}\sum^{29}_{j=0}t_{ji}$ for zero centering,

\begin{equation}
\tilde{\mathbf{T}}_i=(\tilde{t}_{0i},\tilde{t}_{1i},\cdots,\tilde{t}_{29i})=(t_{0i}-\bar{t}_i,t_{1i}-\bar{t}_i,\cdots,t_{29i}-\bar{t}_i)
\end{equation}

The zero-centered PMT arrival time pattern vector for data is

\begin{equation}
\tilde{\mathbf{T}}=(\tilde{t}_{0},\tilde{t}_{1},\cdots,\tilde{t}_{29})=(t_{0}-\bar{t},t_{1}-\bar{t},\cdots,t_{29}-\bar{t})
\end{equation}
where $\bar{t}=\frac{1}{30}\sum^{29}_{j=0}t_{j}$. We constructed corresponding Euclidean distance,

\begin{equation}
d_i=\lvert\tilde{\mathbf{T}}_i-\tilde{\mathbf{T}}\rvert=\sqrt{\sum_{j=0}^{29}\left(\tilde{t}_{ji}-\tilde{t}_{j}\right)^2}
\end{equation}

Then we searched for the $k$ nearest neighbors, where the hyper-parameter $k$ is an arbitrary integer to be chosen later. The reconstructed muon direction $\mathbf{P}$ was calculated by the weighted average of the $k$ nearest neighbors,

\begin{equation}
\mathbf{P} = \dfrac{\sum_{i=1}^k\dfrac{1}{d_i}\mathbf{p}_i}{\sum_{i=1}^k\dfrac{1}{d_i}}
\end{equation}

\begin{figure}[htbp]
\centering
\includegraphics[width=0.4\textwidth]{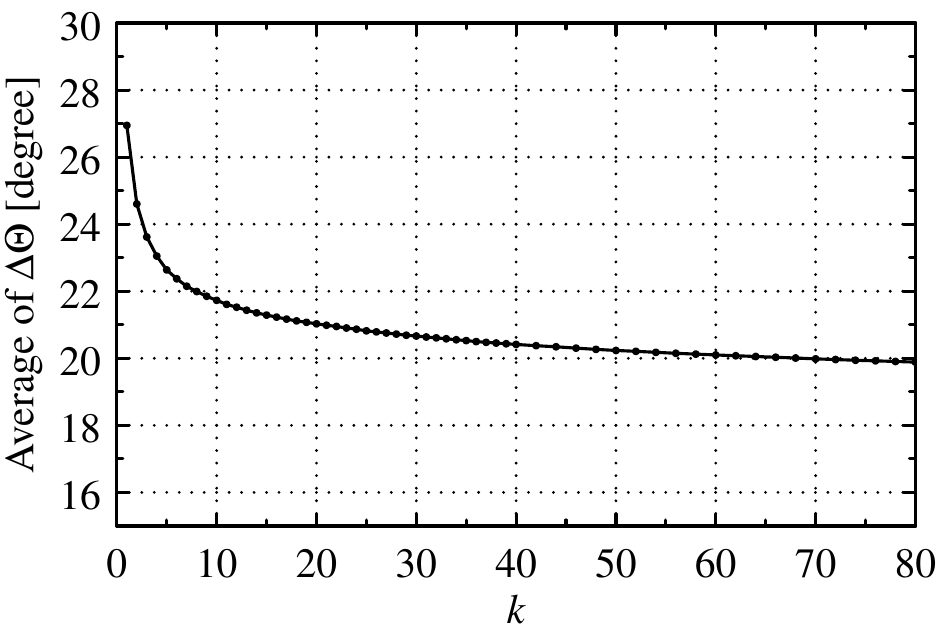}
\caption{The average included angle $\Delta\Theta$ between the truth and reconstructed directions varies with the hyper-parameter $k$.}
\label{fig:hp} 
\end{figure}

\begin{figure}[htbp]
\centering
\includegraphics[width=0.4\textwidth]{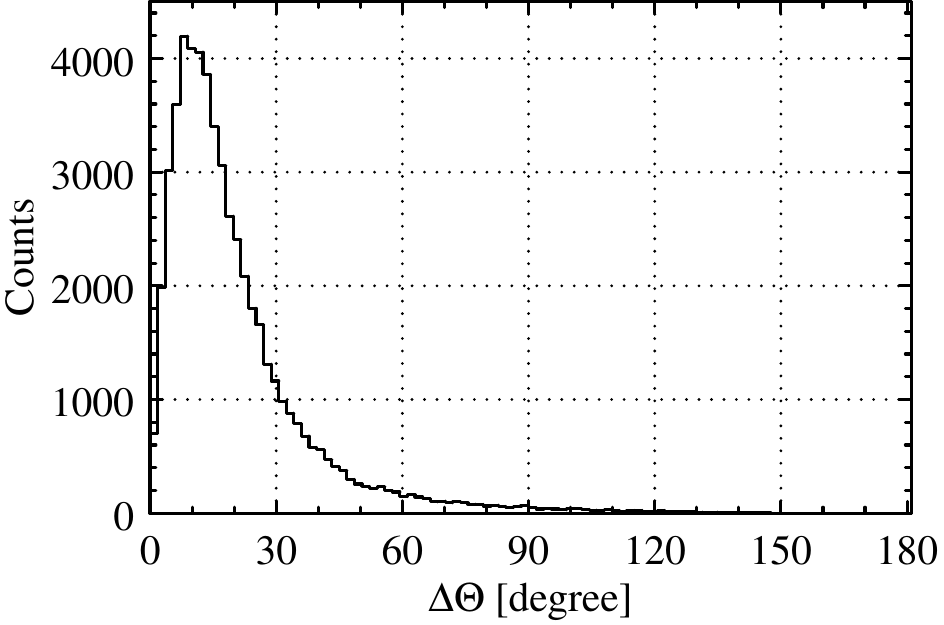}
\caption{The included angle $\Delta\Theta$ between the truth and the reconstructed directions for $k=50$.}
\label{fig:reconperf} 
\end{figure}

We generated a test sample (also uniform muons) to evaluate the reconstruction method's performance and determined the hyper-parameter $k$. The smearing induced by the detector response was included in the test sample to simulate the uncertainty from the electronic hardware and the time calibration. Figure~\ref{fig:hp} shows that the average included angle between the truth and the reconstructed directions $\Delta\Theta$ varies with the hyper-parameter $k$ and becomes stable at $k =50$ and above. Therefore, we chose $k=50$ for the reconstruction. Figure~\ref{fig:reconperf} shows the included angle's distribution with a peak value of 10 degrees and a average of 20 degrees. The long tail was due to the limited time resolution of electronic hardware and PMTs.

\begin{figure}[htbp]
\centering    
\includegraphics[width=0.65\textwidth]{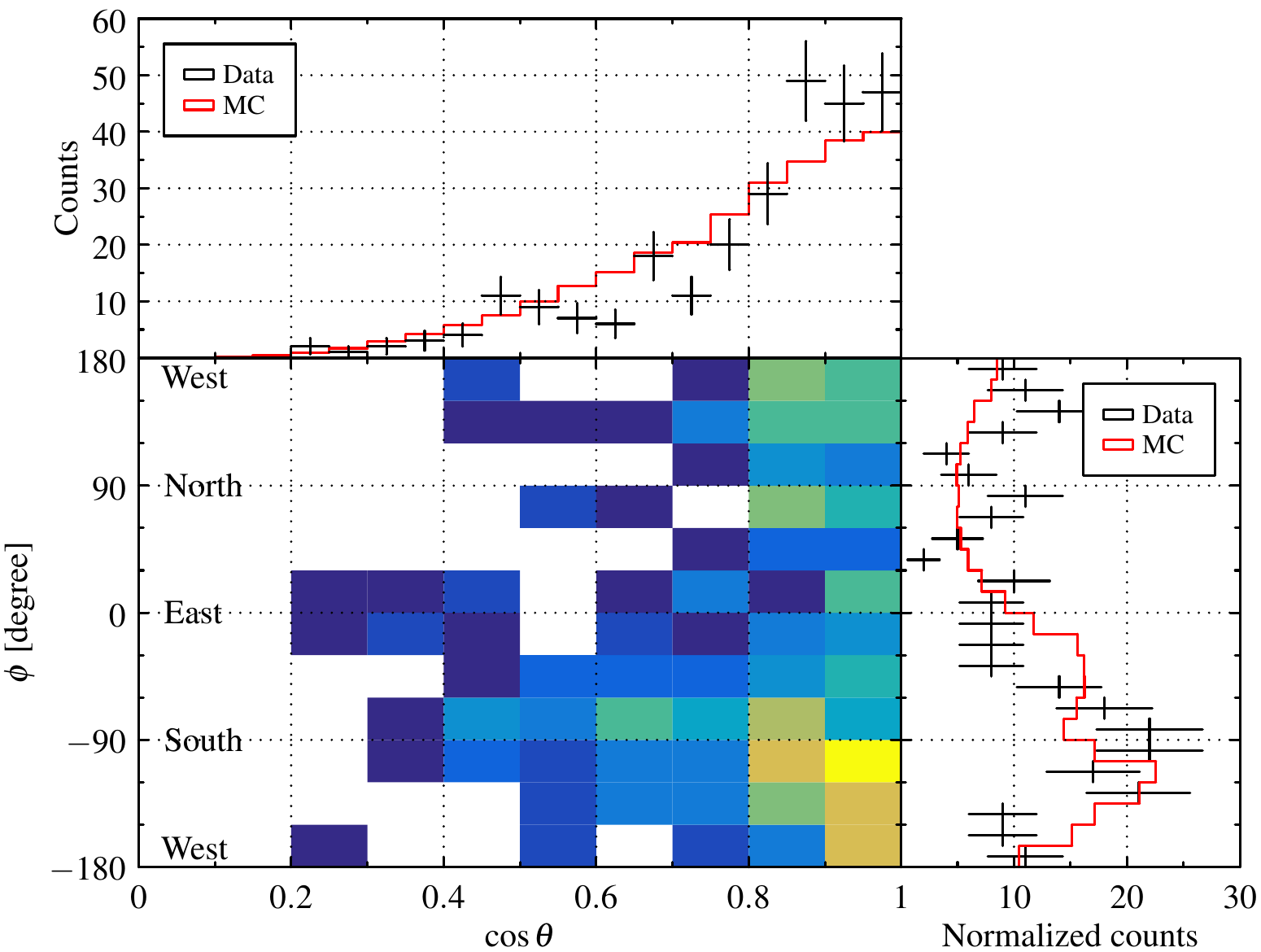}

\caption{The reconstructed $\cos\theta$ and $\phi$ for the selected muon candidates. Also plotted are the one-dimensional projections for these two angles for the data (black ) and the simulation (red).}    
\label{fig:dirfit}
\end{figure}

Figure~\ref{fig:dirfit} shows the $\cos\theta$ and $\phi$ distributions for both the data and the simulation. Both were consistent. The uneven structure observed the $\phi$ distribution indicates the different cosmic-ray leakage due to the mountain structure above CJPL-I. 

\section{Muon flux measurement}
\label{sec:res}

\subsection{Detection efficiency}
\label{sec:detsim}

We defined the overall efficiency $\epsilon$ as the ratio of the number of selected muon candidates $N_\mu$ over the total number of muons going through the experimental hall $N_{\text{total}}$,

\begin{equation}
\epsilon\equiv\frac{N_{\mu}}{N_{\text{total}}}
\end{equation}

\begin{figure}[htbp]
\centering
\includegraphics[width=0.4\textwidth]{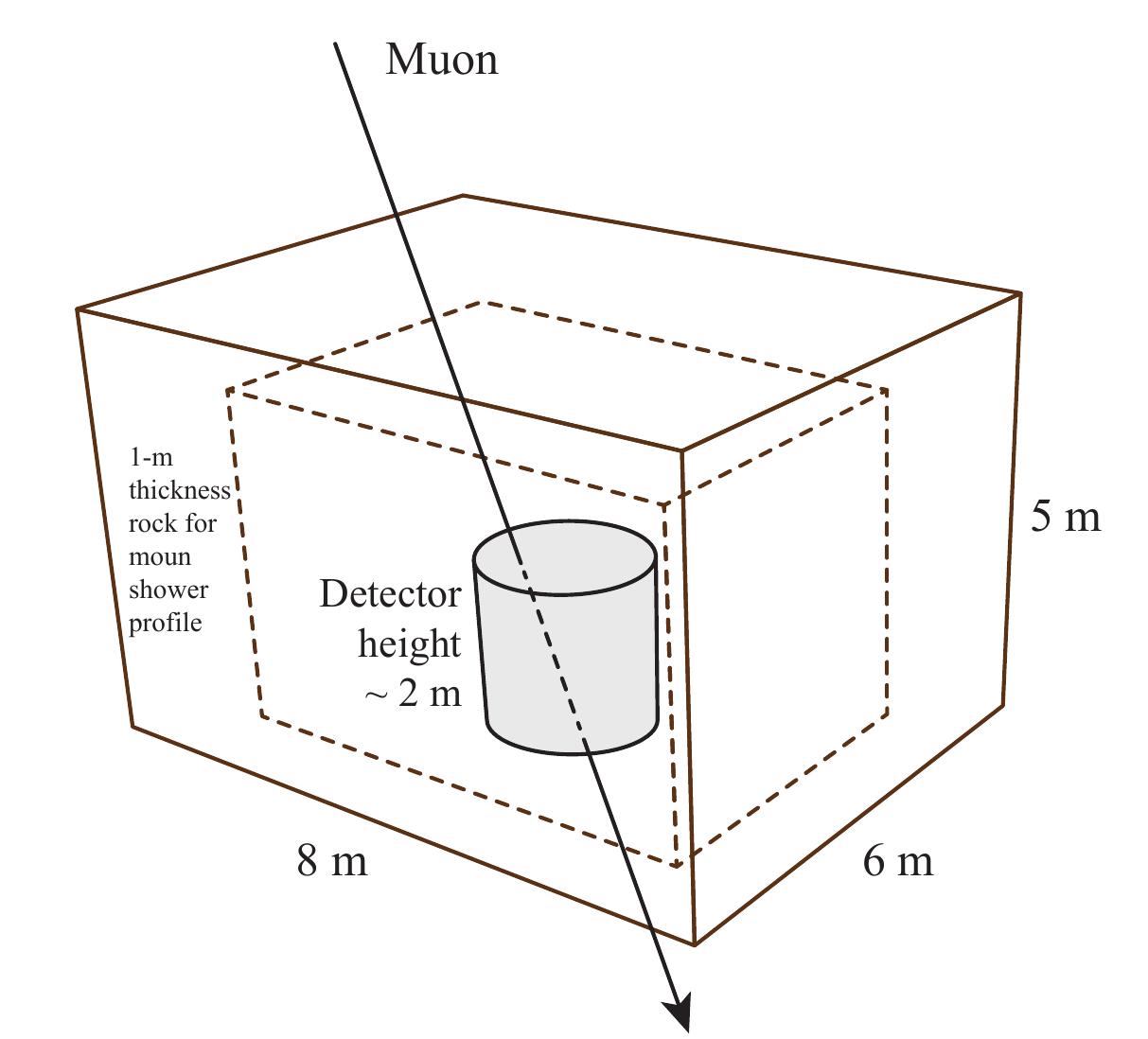}
\caption{Geometry setup in the simulation.}
\label{fig:sd} 
\end{figure}

It is noted that the selected muon candidates contain a small fraction of the muon shower events due to the muon's interaction with the atom of the rock. As a consequence, the incident particles are the parent muons. We had to evaluate this effect in the detection efficiency so we simulated the whole experimental hall. With the underground muon energy spectrum and angular distribution in Section~\ref{sec:mpre}, we performed another Geant4-based simulation to study the detection efficiency. Considering the effect of muon showers in the rock and detector components, we added 1\,m thick rock in the geometry and generated muons flying towards the five surfaces (exclude the bottom surface) of an 8\,m$\times$6\,m$\times$5\,m experimental hall, as shown in Figure~\ref{fig:sd}. 

After decomposed into a geometry factor $\epsilon_g$, a detection efficiency $\epsilon_d$, and a shower factor $\epsilon_s$, the overall efficiency becomes

\begin{equation}
\epsilon = \epsilon_g\cdot\epsilon_d+\epsilon_s,
\end{equation}

\begin{equation}
\epsilon_g=\frac{N_p}{N_{\text{total}}},\quad\epsilon_d=\frac{N_1}{N_p},\quad
\epsilon_s=\frac{N_2}{N_{\text{total}}},
\end{equation}
where $N_p$ is the number of muons passing through the scintillator target, $N_1$ is the number of detected muons passing through the scintillator target, $N_2$ is the number of detected muon shower events, $N_\mu=N_1+N_2$. The simulation result showed that 

\begin{equation}
\epsilon=1.70\%,\quad\epsilon_g=2.02\%,\quad\epsilon_d=82.7\%,\quad\epsilon_s=0.04\%.
\end{equation} 

The detection efficiency $\epsilon_d=82.7\%$ indicated that the efficiency loss in the event selection is less than 20\%.

\begin{figure}[htbp]
\centering
\includegraphics[width=0.35\textwidth]{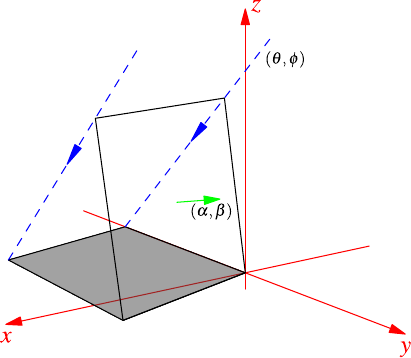}
\caption{Projection area (gray) of the surface $(\alpha,\beta)$ for the muon direction $(\theta,\phi)$.}
\label{fig:pa} 
\end{figure}

As shown in Figure~\ref{fig:pa}, for a surface $S$ with normal direction $(\alpha_\text{zenith},\beta_\text{azimuth})$ and muon direction $(\theta, \phi)$, the horizontal projection area $S_\text{p}$ is given by,

\begin{equation}
S_\text{p}(\theta, \phi)=\left|\sin \alpha  \tan \theta  \cos (\beta -\phi )+\cos \alpha \right|S
\end{equation}

We defined $S_{i}$ as the projection area of $i$-th surface, which is an integral for the normalized incoming muon spectrum $f(E_k, \theta,\phi)$ obtained from the simulation result of Section~\ref{sec:mpre},

\begin{equation}
S_i=\int S_{\text{p}}(\theta, \phi) f(E_k, \theta,\phi)\,\mathrm{d}(\cos\theta)\mathrm{d}\phi
\end{equation}

The experimental hall's projection area $S_{\text{total}}$ is a sum of the five surface, 

\begin{equation}
S_{\text{total}}=\sum_{i=1}^5S_i=78.7\,\textrm{m}^2.
\end{equation}

\subsection{Muon flux result}
\label{sec:mr}
The muon flux $\phi_\mu$ was calculated by

\begin{equation}
\phi_\mu=\frac{N_{\text{total}}}{T\cdot S_{\text{total}}}=\frac{N_\mu/\epsilon}{T\cdot S_{\text{total}}}
\end{equation}
where $T$ is the live time, $N_\mu$ is the number of muon candidates. By defining $S\equiv\epsilon S_{\text{total}}$ as the active area, we simplified the 1-ton prototype detector's flux calculation,

\begin{equation}
\phi_\mu=\frac{N_\mu}{T\cdot S}=3.53\times10^{-10}\,\text{cm}^{-2}\text{s}^{-1}
\end{equation}

The simulation result showed that the active area $S=1.34$\,m$^2$ is close to the cross-section of the liquid scintillator sphere of the detector, which is 1.31\,m$^2$.

\subsection{Systematic uncertainties}
\label{sec:unc}

Table~\ref{tab:su} summarizes the systematic uncertainties, which mainly come from two parts: (1) the PE number calculation (energy scale)
in the data and (2) the efficiency calculation in the Monte-Carlo simulation. A quadrature sum of the individual components gives the total systematic uncertainty. 

\begin{table}[htbp]
\centering
\caption{Summary of uncertainties for the muon flux measurement. }
\label{tab:su}
\begin{tabular}{lll}
\toprule
\noalign{\smallskip}
\multirow{2}{*}{\makecell[c]{Source}} &  Parameter  & Flux measurement \\ 
 & uncertainty & uncertainty \\
\noalign{\smallskip}
\midrule
Energy scale & $\pm2.0\%$ & $\pm0.6\%$ \\
Efficiency calculation \\
\qquad PE yield & $\pm1.6\%$ & $\pm0.5\%$\\
\qquad Acrylic vessel radius & $\pm0.5$\,cm & $\pm1.6\%$\\
\qquad Lead shielding thickness & $\pm5\,$cm & $\pm0.6\%^*$\\
\qquad Rock thickness for muon shower profile & $\pm50\,$cm & $\pm0.8\%^*$\\
\qquad Muon spectra & - & $\pm0.7\%^*$ \\
Total systematic & - & $\pm2.2\%$\\
Statistics & - & $\pm6.2\%$\\ 
\bottomrule
\end{tabular}
\footnotesize{\\$^*$ Dominated by the statistics uncertainty of Monte-Carlo.}\\
\end{table}

The conversion from the charge to the number of PEs was through a PMT gain factor. A run-by-run PMT calibration corrected the gain drift and introduced a 2.0\% systematic uncertainty for the 6,000 p.e. cut, corresponding to a 0.6\% efficiency variation in flux measurement.

The uncertainty of efficiency calculation came from the parameters in the simulation's input. We compared the data and simulation's PE distribution and tuned the scintillation light yield to ensure consistency between the data and simulation. The evaluation of systematic uncertainty for the consistency was from studying the Person's $\chi^2$ as given below,

\begin{equation}
\chi^2=\sum_{i=1}^{\text{nbins}}\frac{(n_{i}^{\text{data}}-n_{i}^{\text{sim}})^2}{n_i^{\text{data}}}
\end{equation}
where $n_{i}^{\text{data}}$ is the count in the $i$-th bin for the data distribution, $n_i^{\text{sim}}$ is the count in the $i$-th bin for the simulation distribution.
The systematic uncertainty of PE yield in the simulation was set to the variation at $\chi^2_{\min}+1$.

Two hemispheres glued the acrylic vessel and the machining accuracy was estimated to 5\,mm according to the international tolerance grade, which contributed a 1.6\% systematic uncertainty for the efficiency calculation.

The muon shower in the rock and lead shielding also contributed to the global efficiency. We placed 1\,m depth of rock in the simulation. To verify whether the depth is enough, we added/subtracted 0.5\,m rock in different simulations to observe the variation and found that the global efficiency was not sensitive to rock depth. The lead wall thickness was not uniform due to the different lead brick arrangement. We changed the thickness by $\pm5$\,cm, a typical size of a lead brick, in the simulations, and found little variation in the global efficiency. Limited by the statistical uncertainty of Monte-Carlo, the above studies gave 0.6\% and 0.8\% systematic uncertainty for the muon shower effect. 

We scanned different muon spectra (energy and angular distribution) in Section~\ref{sec:mpre} in the detection efficiency simulation. Thanks to the detector's spherical symmetry, the uncertainty from the muon spectra was also small and dominated by the statistical uncertainty of Monte-Carlo.

\subsection{Discussion}

\begin{figure}[htbp]
\centering
\includegraphics[width=0.7\textwidth]{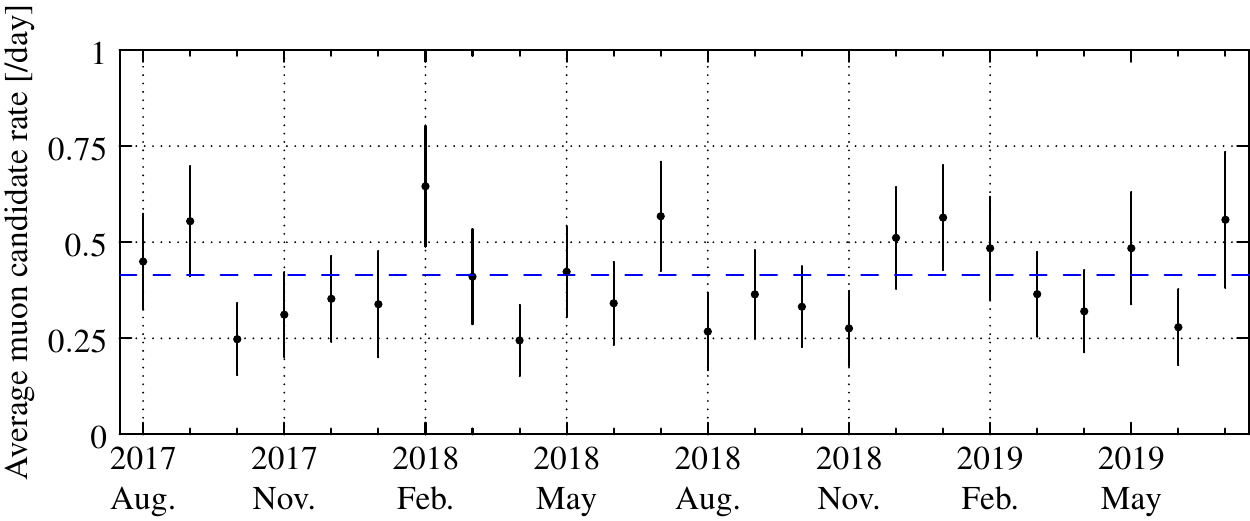}
\caption{Cosmic-ray muon rate measured by the 1-ton prototype at CJPL-I, as a function of time. The data are shown in monthly bins.}
\label{fig:muonmon} 
\end{figure}

The monthly average muon candidate rate shows a constant in Figure~\ref{fig:muonmon}. The seasonal modulation of muon flux is unobservable because the statistics uncertainty is much larger than the seasonal variation ($\sim1.3\%$ in Ref.~\cite{borexinom}). The total measured cosmic-ray muon flux was $(3.53\pm0.22_{\text{stat.}}\pm0.07_{\text{sys.}})\times10^{-10}$\,cm$^{-2}$s$^{-1}$. 
The vertical intensity $I$ was calculated to be $(2.0\pm0.3_{\text{stat.}})\times10^{-10}$\,cm$^{-2}$s$^{-1}$sr$^{-1}$ using the 47 muon candidates ($N_I$) reconstructed in $0.95<\cos\theta<1$,

\begin{equation}
I=\frac{N_I}{T\cdot S\cdot\Omega}
\end{equation}
where $T$ and $S$ are the live time and active area described in Sec.~\ref{sec:mr}, $\Omega=0.314$ is the solid angle for $0.95<\cos\theta<1$. We also studied the flux variation as a function of horizontal location at CJPL-I by a Geant4-based simulation. The result indicated that the variation should be less than 2.3\% along the west-east direction within a variation of 100\,m.  

\begin{figure}[htbp]
\centering
\includegraphics[width=0.5\textwidth]{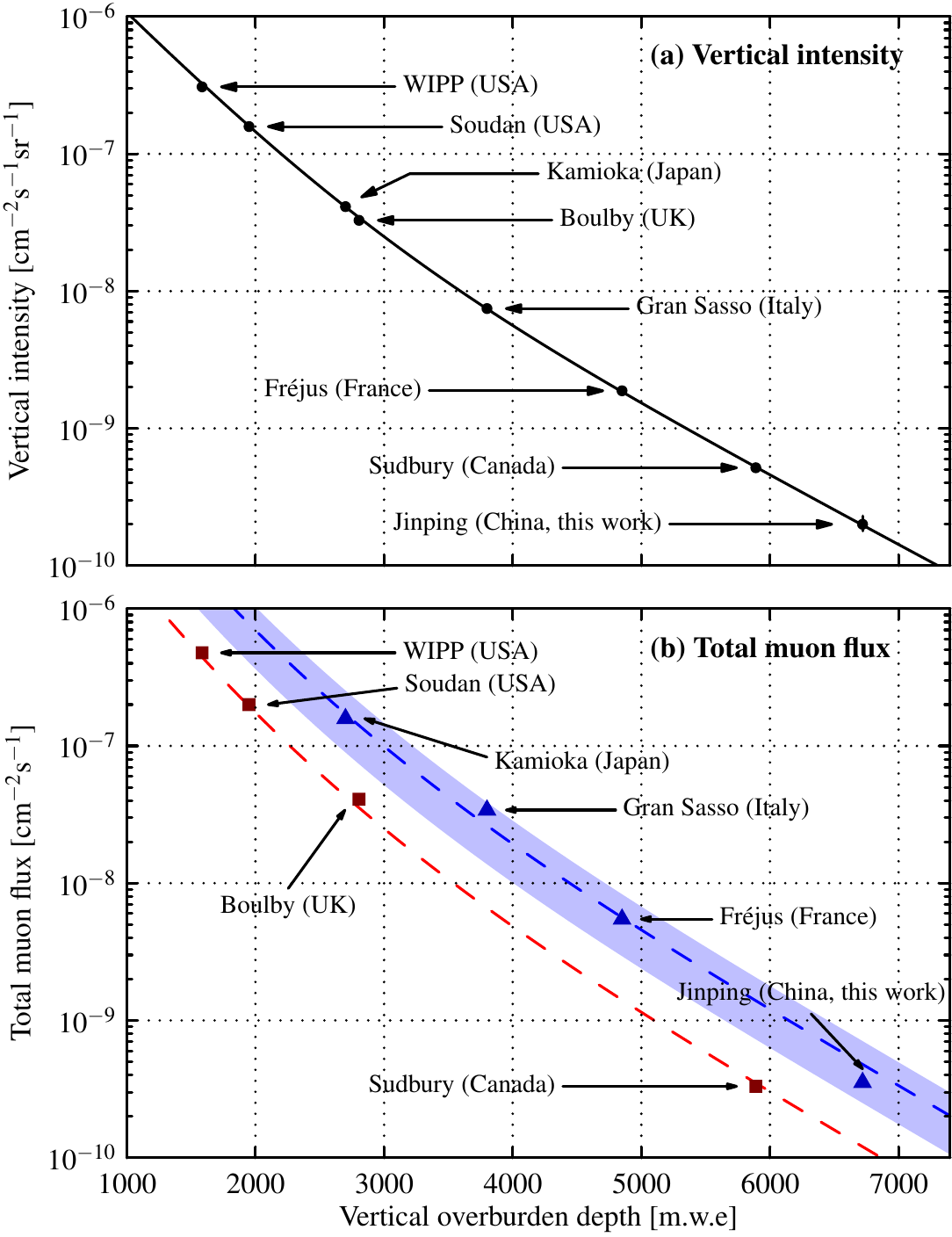}
\caption{Measurements of the vertical intensity (a) and total muon flux (b) at different underground sites. The red dashed line in (b) is the empirical formula for the down mine shaft case. The blue dashed line and shade in (b) are the fitting result and uncertainty for the factor $F$.}
\label{fig:lab} 
\end{figure}

Figure~\ref{fig:lab}(a) shows the vertical intensity of muons at WIPP\cite{wipp}, Soudan\cite{soudan}, Boulby\cite{boulby}, Sudbury\cite{snom}, Kamioka\cite{kamioka}, Gran Sasso\cite{borexinom}, Fréjus\cite{frem}, and Jinping as a function of vertical overburden. Also plotted is the prediction by a parametrized formula, given by Ref.~\cite{Mei:2005gm},

\begin{equation}
I(h)=I_1\ue^{-h/\lambda_1}+I_2\ue^{-h/\lambda_2}
\label{eq:vm}
\end{equation}
where $I(h)$ is the differential muon intensity corresponding to the slant depth $h$, $I_1, I_2, \lambda_1, \lambda_2$ are parameters in Ref.~\cite{Mei:2005gm}. The measurement result in this work is consistent with Eq.~(\ref{eq:vm}).

Figure~\ref{fig:lab}(b) summarized the total muon flux measured at different underground sites. WIPP, Soudan, Boulby, and Sudbury are the labs situated down mine shafts, while Kamioka, Gran Sasso, Fréjus, and Jinping are below mountains. The underground muon flux $\phi$ is a complicated integral over the muon spectrum $f(E_k,\theta,\phi)$ and slant depth (or muon track length) $d$,

\begin{equation}
\phi=\int f(E_k,\theta,\phi)\cdot p(E_k,d)\,\ud E_k\ud\theta\ud\phi
\label{eq:fluxint}
\end{equation}
where $p(E_k,d)$ is the survival probability at slant depth $d$. For the down mine shaft (flat overburden) case, we have $d=h/\cos\theta$. We simulated the muon flux at 50 different depths for the flat overburden case by Geant4. For simplicity, we fitted these simulation points with a series expansion of Eq.~(\ref{eq:fluxint}),

\begin{equation}
\phi_0(h)=\exp\left(a_0+a_1h+a_2h^2+a_3h^3+a_4h^4\right)\,\text{cm}^{-2}\text{s}^{-1}
\label{eq:muonh}
\end{equation}
where $\phi_0(h)$ is the muon flux at vertical overburden depths $h$ (in km.w.e), $a_0=-10.147$, $a_1=-3.385\,\text{km}^{-1}$, $a_2=0.404\,\text{km}^{-2}$, $a_3=-0.0344\,\text{km}^{-3}$, $a_4=0.00111\,\text{km}^{-4}$. The differences between the empirical formula and data were $-6.9\%$ (WIPP), $-2.6\%$ (Soudan), $-13.3\%$ (Boulby), and $5.7\%$ (Sudbury). The red dashed line in Figure~\ref{fig:lab}(b) plots the fitting result.

The total muon flux of a lab situated below a mountain, $\phi_1(h)$, can be scaled by a factor $F$ to $\phi_0(h)$, 

\begin{equation}
\phi_1(h)=F\cdot\phi_0(h)
\label{eq:phi1}
\end{equation}
usually $F>1$ because the mountain case has less rock shielding and leads to a more considerable muon flux. The factors $F$ were 3.7 (Kamioka), 5.2 (Gran Sasso), 3.9 (Fréjus) and 2.9 (Jinping). We assumed that the mountains on the Earth have similar topography and elemental compositions so that the factors for different locations would not vary too much. We fitted these four labs using the empirical formula in Eq.~(\ref{eq:phi1}) with an uncertainty assigned so that $\chi^2/$ndf is one. The fitting result $F=(4.0\pm1.9)$ and this uncertainty accounted for the variation of mountain topography profiles. The blue dashed line in Figure~\ref{fig:lab}(b) illustrates the fitting result. Using $\phi_0$ and factor $F$, anyone can easily get a rough estimate from the vertical overburden without doing any simulation or measurement.

\section{Summary}

We studied the cosmic-ray muons at CJPL-I using the 1-ton prototype of the Jinping Neutrino Experiment. This study determined the muon flux to be $(3.53\pm0.22_{\text{stat.}}\pm0.07_{\text{sys.}})\times10^{-10}$\,cm$^{-2}$s$^{-1}$. The zenith and azimuth angle distributions show that cosmic-ray leakage is due to the mountain topography profile as expected. A survey of muon fluxes at different locations of laboratory situated under mountains and below mine shafts indicated that the former is generally a factor of $(4.0\pm1.9)$ larger than the latter for the same vertical overburden. This study provides a reference for passive and active shielding designs for future underground neutrino experiments.

\section*{Acknowledgements}
This work was supported in part by, the National Natural Science Foundation of China (No.~11620101004 and 11475093), the Key Laboratory of Particle \& Radiation Imaging (Tsinghua University), the CAS Center for Excellence in Particle Physics (CCEPP), and Guangdong Basic and Applied Basic Research Foundation under Grant No. 2019A1515012216.
Portion of this work performed at Brookhaven National Laboratory is supported in part by the United States Department of Energy under contract DE-SC0012704. We acknowledge Orrin Science Technology, Jingyifan Co., Ltd, and Donchamp Acrylic Co., Ltd, in their efforts on the engineering design and fabrication of the stainless steel and acrylic vessels. Many thanks to the CJPL administration and the Yalong River Hydropower Development Co., Ltd. for the logistics and supports.

\bibliography{main}

\end{document}